\documentclass{emulateapj}
\usepackage{amsmath,color}
\usepackage{array}
\usepackage{float}
\usepackage{graphicx}
\usepackage{subfigure}
\usepackage{natbib}

\begin{document}
\title{The color-magnitude distribution of Hilda asteroids: Comparison with Jupiter Trojans}
\author{Ian Wong and Michael E. Brown}
\affil{Division of Geological and Planetary Sciences, California Institute of Technology,
Pasadena, CA 91125, USA; iwong@caltech.edu}

\begin{abstract}
Current models of Solar System evolution posit that the asteroid populations in resonance with Jupiter are comprised of objects scattered inward from the outer Solar System during a period of dynamical instability. In this paper, we present a new analysis of the absolute magnitude and optical color distribution of Hilda asteroids, which lie in the 3:2 mean motion resonance with Jupiter, with the goal of comparing the bulk properties with previously published results from an analogous study of Jupiter Trojans. We report an updated power law fit of the Hilda magnitude distribution through $H=14$. Using photometric data listed in the Sloan Moving Object Catalog, we confirm the previously-reported strong bimodality in visible spectral slope distribution, indicative of two sub-populations with differing surface compositions. When considering collisional families separately, we find that collisional fragments follow a unimodal color distribution with spectral slope values consistent with the bluer of the two sub-populations. The color distributions of Hildas and Trojans are comparable and consistent with a scenario in which the color bimodality in both populations developed prior to emplacement into their present-day locations. We propose that the shallower magnitude distribution of the Hildas is a result of an initially much larger Hilda population, which was subsequently depleted as smaller bodies were preferentially ejected from the narrow 3:2 resonance via collisions. Altogether, these observations provide a strong case supporting a common origin for Hildas and Trojans as predicted by current dynamical instability theories of Solar System evolution.
\end{abstract}

\section{Introduction}

Over the past few decades, the classical picture of Solar System formation and evolution, in which planets formed and smoothly migrated to their present-day locations within the protoplanetary disk, has been beset by significant challenges. The unexpectedly high eccentricities and inclinations of the giant planets, the dynamically excited orbital distribution of the Kuiper Belt, and the irregular satellites of Jupiter and Saturn are among an increasing body of observations that point toward a chaotic restructuring of the Solar System orbital architecture after the dispersal of the protoplanetary disk. 

Current theories of Solar System evolution describe a scenario in which Jupiter and Saturn crossed a mean-motion resonance, setting off a period of dynamical instability throughout the middle and outer Solar System \citep[e.g.,][]{morbidelli}. Simulations have shown that the primordial minor body populations in resonance with Jupiter (Hildas and Jupiter Trojans) were first emptied during this turbulent episode and then replaced almost exclusively with planetesimals scattered inward from the region beyond the ice giants \citep[][]{gomes,roig2}. The major implication of these models is that Kuiper Belt objects, Trojans, and Hildas all originated within a single progenitor population in the outer Solar System and should therefore be largely identical. By comparing the observable properties of Hildas and Trojans, one can evaluate their similarities and/or differences and thereby empirically test current dynamical instability models.

Recent progress in our understanding of Hildas and Trojans has already uncovered many notable similarities. Objects in both populations share the general characteristics of flat, featureless optical and near-infrared spectra with reddish colors \citep[e.g.,][]{Hildaspectra,dotto,fornasier,marsset} and similar, very low albedos \citep[e.g.,][]{fernandez2003,fernandez,Hildaalbedo}. In addition, both Hildas and Trojans are notable in having a bimodal color distribution. Analyses of spectral slopes derived from photometry contained in the Sloan Digital Sky Survey Moving Object Catalog (SDSS-MOC) for both Hildas \citep{gilhutton} and Trojans \citep[e.g.,][]{roig,wong} demonstrate a clear bifurcation in the optical color distribution and indicate the presence of two classes of objects within the Hildas and Trojans. The bimodality in optical color is supported by bimodality in the infrared reflectivity distribution measured by WISE and NEOWISE for both Hildas \citep{grav2} and Trojans \citep{grav}.

The most direct way of comparing two populations is by studying their bulk properties, namely, the absolute magnitude distribution and the color distribution. For a population with a narrow range of albedos, such as the Hildas and Trojans, the magnitude distribution is a good proxy for the size distribution and contains information about both the formation environment and the subsequent collisional evolution of the population. The color distribution reveals the diversity of surface types and also provides constraints on models of the composition and origin of objects within the population. 

In \citet{wong}, we carried out an in-depth study of the color-magnitude distribution of Trojans. In this paper, we present an analogous study for Hildas, in order to obtain a point of reference for comparing the two populations. We report fits to the total Hilda magnitude distribution and provide a detailed analysis of the updated color distribution, as derived from the newest fourth release of the Sloan Moving Object Catalog (SDSS-MOC4). Special attention is given to exploring the properties of individual collisional families. The results of our Hilda analysis are compared with our previously published Trojan results and discussed in relation to collisional evolution, surface composition, and dynamical considerations within the framework of recent dynamical instability models of Solar System evolution

\section{Data and analysis}

In this section, we present our analysis of the absolute magnitude and optical color distributions of the Hilda asteroids. The methods used are mostly identical to those described in detail in our previously published analysis of Jupiter Trojans \citep[see][and references therein]{wong}.

In selecting for Hilda asteroids, we have applied the following constraints in orbital parameter space, which are used by the IAU Minor Planet Center (MPC) in their definition of Hildas: $3.7 \le a \le 4.1$~AU, $e \le 0.3$, and $i \le 20^{\circ}$. Querying the MPC database with these criteria results in a total count of 3801 Hildas (as of October 2016). Using a less stringent criterion \citep[e.g., extending the maximum values of $(a,e,i)$ to $(4.2~\mathrm{AU},0.4,30^{\circ})$, as in][]{grav2} does not appreciably increase the overall number of Hildas and does not significantly affect the results of our analysis. 

\begin{figure}[t!]
\begin{center}
\includegraphics[width=9cm]{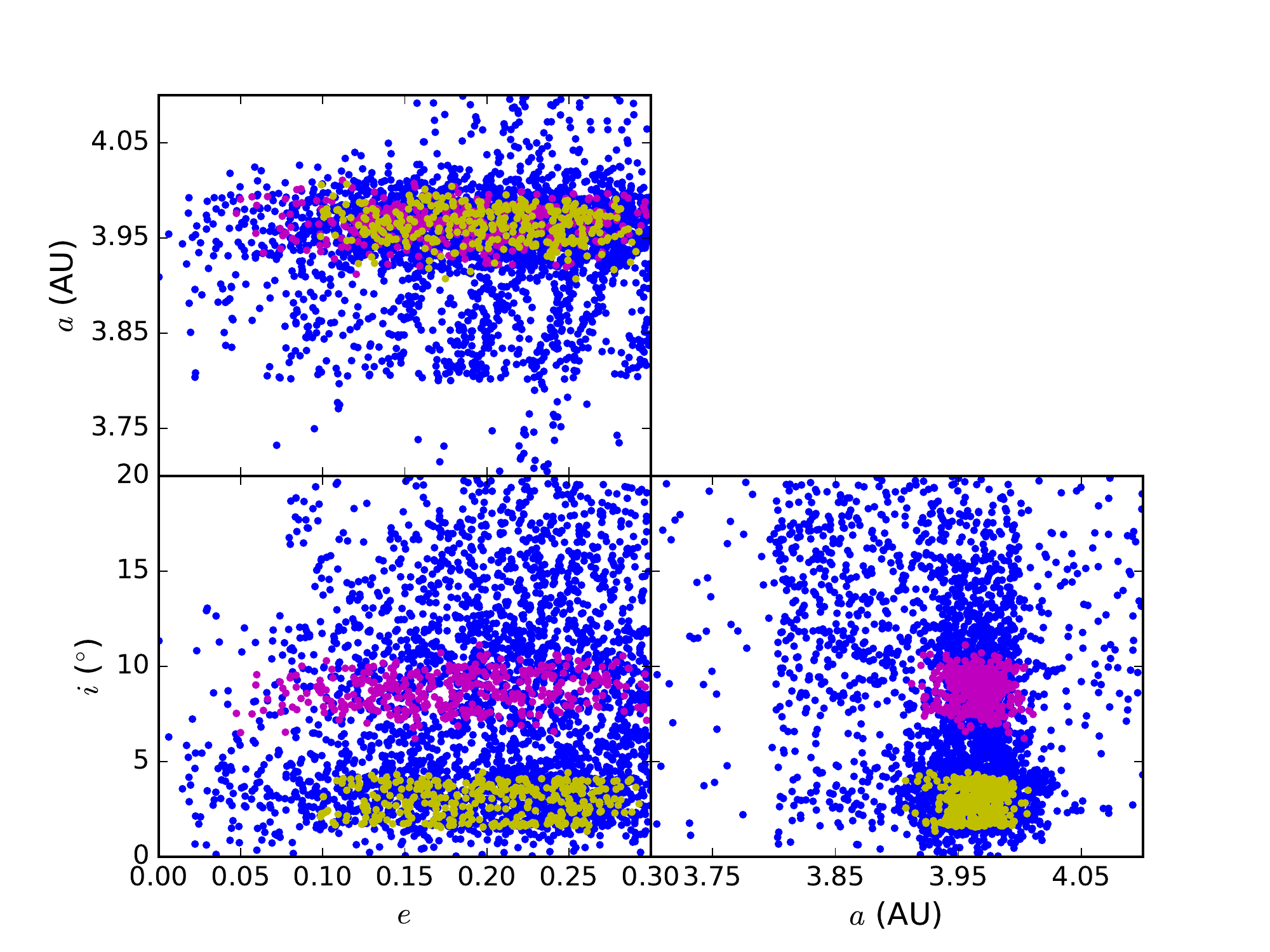}
\end{center}
\caption{Distribution of the 3801 objects in our Hilda dataset, plotted in the space of semi-major axis ($a$), eccentricity ($e$), and inclination ($i$). Objects belonging to the Hilda and Schubart collisional families are denoted by magenta and yellow dots, respectively; background Hildas are denoted by blue dots.} \label{distall}
\end{figure}

Studying the color-magnitude distribution of Hilda collisional families is of particular relevance in our understanding of the composition and evolution of the population as a whole. There exist two major collisional families within the Hildas --- the Hilda and Schubart families. We have created lists of family members using the tabulated results in \citet{nesvorny}, which identify 385 members of the Hilda family and 350 members of the Schubart family. The distribution of Hildas in $(a,e,i)$ space is illustrated in Figure~\ref{distall}, with the location of the two collisional families highlighted.

\subsection{Magnitude distributions}
\begin{figure}[t!]
\begin{center}
\includegraphics[width=9cm]{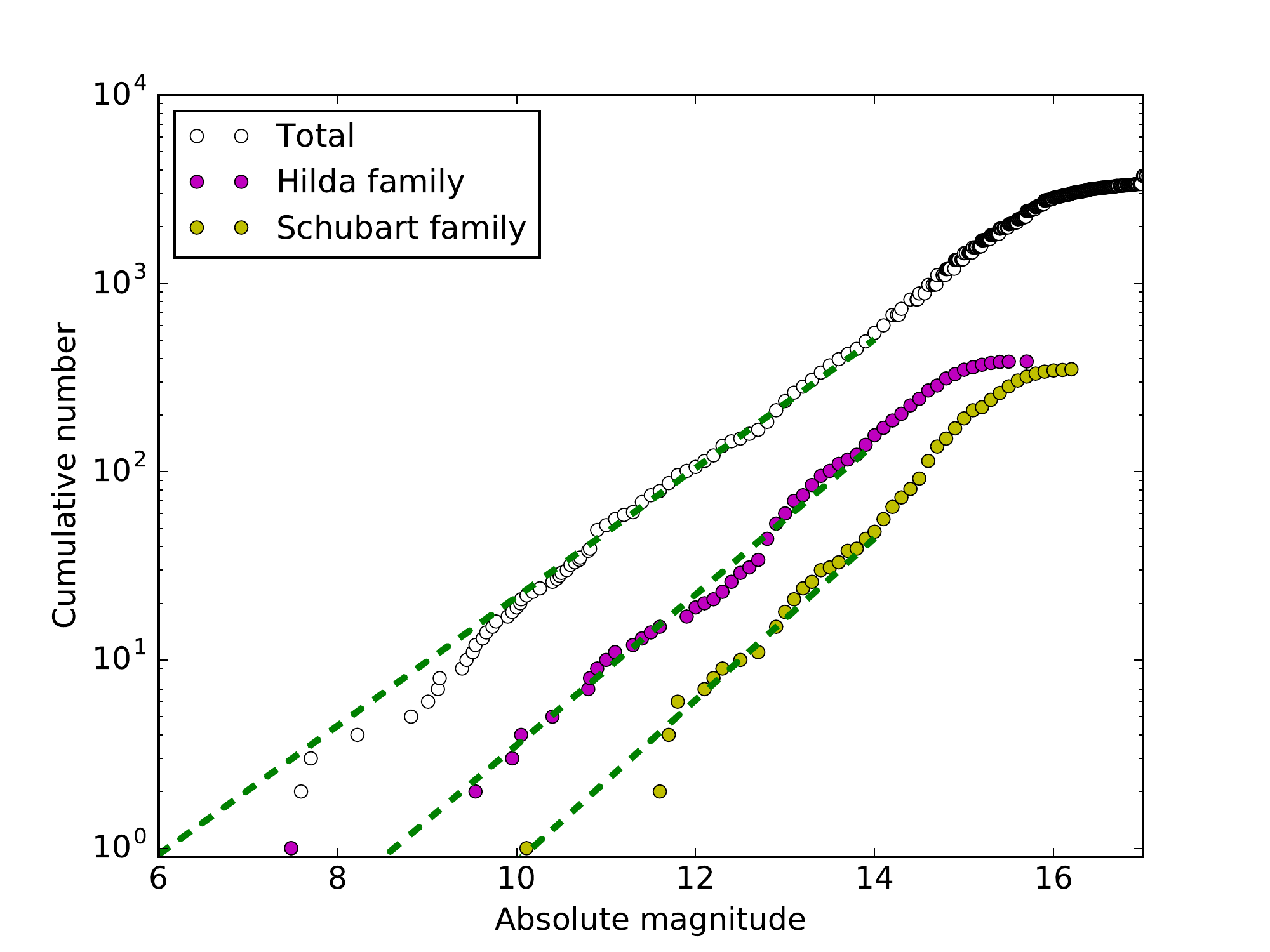}
\end{center}
\caption{Cumulative absolute magnitude distributions of the total Hilda population (white), as well as the Hilda and Schubart collisional families individually (magenta and yellow, respectively). The best-fit power law curves describing the distributions are overplotted (dashed green lines).} \label{Hildamag}
\end{figure}

For each object, we set the absolute magnitude to the value listed in the Asteroid Orbital Elements Databse. The cumulative absolute magnitude distribution of the total Hilda population is shown in Figure~\ref{Hildamag}. The magnitude distribution has the characteristic shape seen in many minor body populations, with a steeper slope at large sizes transitioning to a shallower slope at intermediate sizes. 

The gentle rollover at $H\sim16$ reflects the onset of incompleteness in the Hilda dataset. In our analysis of Trojans, we determined the completeness limit  of the MPC Trojan dataset to be $H \sim 11.3$ and were able to correct the shape of the magnitude distribution for incompleteness at fainter magnitudes by utilizing the deeper SDSS-MOC4 dataset, which we calculated to be complete for Trojans through $H=12.3$. In the case of Hildas, however, the onset of incompleteness in the SDSS-MOC4 dataset occurs at a brighter magnitude ($H\sim14$) than the MPC dataset, and as such, we are unable to correct for incompleteness in the total magnitude distribution. In this paper, we have chosen a conservative upper limit for our analysis at $H=14$. Varying this limit by 0.5~mag in either direction does not significantly affect the distribution fits.

We fit the total differential magnitude distribution, $\Sigma(H)$, to a single power law of the form
\begin{equation}\label{distribution2}\Sigma(\alpha_{1},H_{0}|H) = 
10^{\alpha(H-H_{0})},\end{equation}
where $\alpha$ is the slope of the distribution, and $H_{0}$ is the threshold magnitude used to properly normalize the distribution to fit the data.

The best-fit parameter values and $1\sigma$ uncertainties were computed using a Markov Chain Monte Carlo (MCMC) ensemble sampler. For the total magnitude distribution, the best-fit parameter values are $\alpha=0.34^{+0.02}_{-0.01}$ and $H_{0}=6.42\pm 0.29$. We also experimented with fitting the total magnitude distribution with a four-parameter broken power law \citep[e.g.,][]{wong,wong2}; however, the addition of a second power law slope is strongly disfavored by the Bayesian Information Criterion ($\Delta$BIC = 8.4; BIC$\equiv -2\log(L)+k\log(n)$, where $L$ is the likelihood of the best-fit solution, $k$ is the number of free parameters, and $n$ is the number of data points).

We also fit the magnitude distributions of the Hilda and Schubart collisional families, which are plotted in Figure~\ref{Hildamag}. The best-fit parameters are $\alpha=0.40^{+0.04}_{-0.03}$ and $H_{0}=8.77^{+0.36}_{-0.37}$ for the Hilda family, and $\alpha=0.43^{+0.07}_{-0.03}$ and $H_{0}=10.23^{+0.39}_{-0.50}$ for the Schubart family. The slopes of the Hilda and Schubart collisional family magnitude distributions are steeper than the overall population (at the $1.7\sigma$ and $2.5\sigma$ levels, respectively), and are consistent with the range of power law slopes derived from numerical simulations of asteroid fragmentation \citep[e.g., $\alpha=$0.44--0.54 in][]{jutzi}.

For each power law fit, we sampled the best-fit distribution to create a model magnitude distribution of the same size as the respective population within the magnitude range under consideration ($H<14$). We carried out a two-sample Anderson-Darling test, which evaluates the null hypothesis that the model distribution and the data are drawn from the same underlying distribution. In all cases, we could not reject the null hypothesis at a confidence level greater than 50\%, demonstrating that the model distribution fits are a statistically good match to the data.

\subsection{Color distribution and sub-populations}
The SDSS-MOC4 lists photometric measurements of minor bodies in the \textit{u, g, r, i, z} bands. We queried the database for Hildas and identified 275 objects that were observed by the Sloan survey. Following the methods of \citet{roig} and \citet{wong}, we corrected the listed apparent magnitudes for solar colors and derived relative reflectance fluxes (normalized to $1$ in \textit{r} band), discarding observations in which any of the band fluxes had a relative error greater than 10\%. For each observation, the spectral slope $S$ was computed from an error-weighted linear least-squares fit to the fluxes in the \textit{g, r, i, z} bands. The \textit{u} band flux was not used in fitting since the flux at those wavelengths typically deviates significantly from the linear trend in the spectrum at longer wavelengths \citep{roig}. For objects with multiple observations, we calculated the weighted average spectral slope.

\begin{figure}[t!]
\begin{center}
\includegraphics[width=9cm]{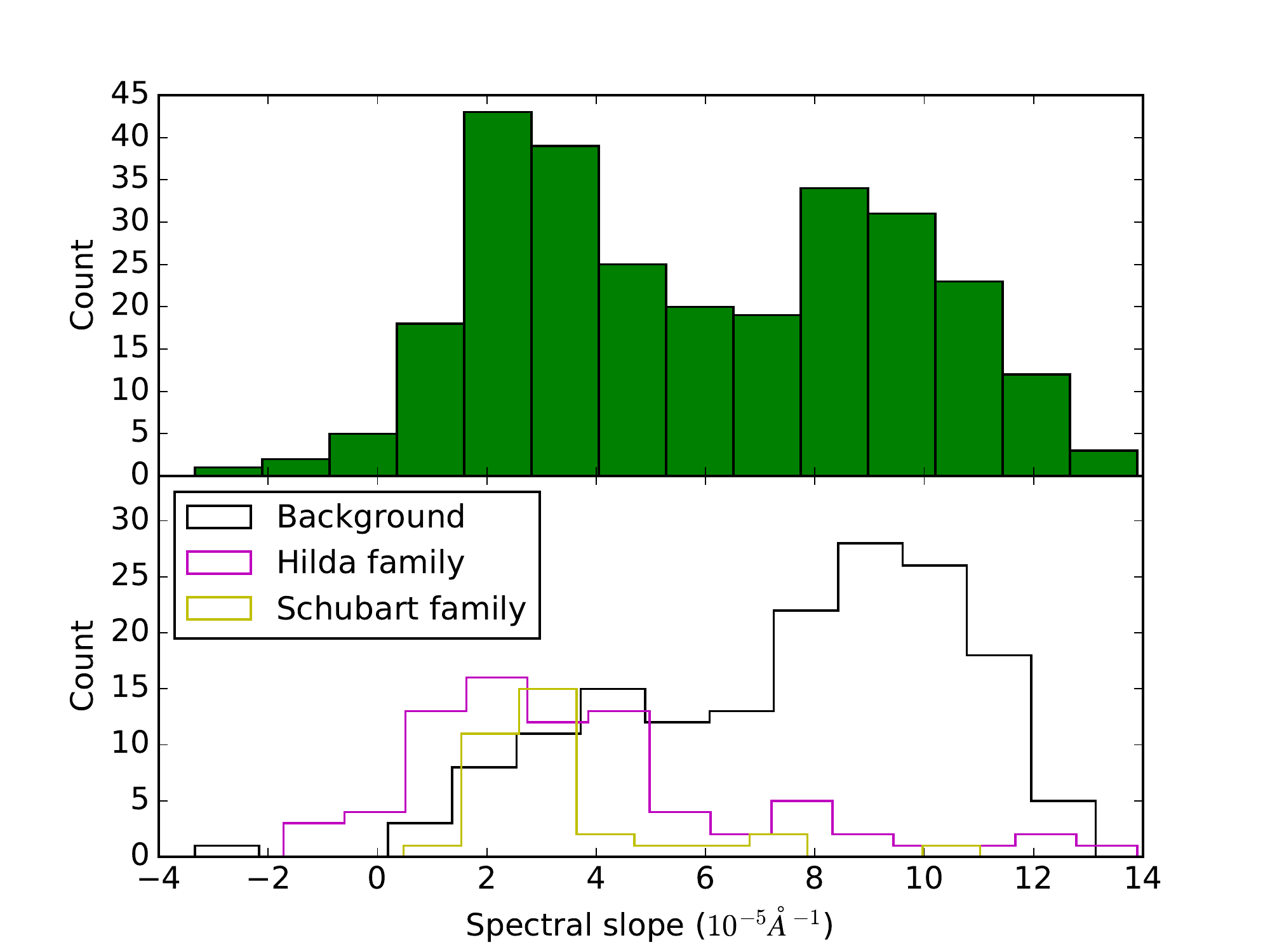}
\end{center}
\caption{Top panel: the overall spectral slope distribution of Hildas, as derived from SDSS-MOC4 photometry, demonstrating a robust color bimodality that divides the population into less-red and red objects. Bottom panel: the spectral slope distributions for Hilda and Schubart family members, as well as background non-family members. Note that the background color distribution is bimodal, while the individual collisional family color distributions are both unimodal.} \label{Hildacol}
\end{figure}

The spectral slope distribution of Hildas is shown in Figure~\ref{Hildacol}, where a clear bimodality is evident, as was first reported in \citet{gilhutton}. The earlier study used the previous, third release of the Moving Object Catalog (SDSS-MOC3) and identified 122 Hildas in the dataset. The latest release more than doubled the number of Hildas with photometric measurements. To quantitatively assess the significance of the bimodality, we fit single and double Gaussian models to the color distribution and found that the two-peaked model is very strongly favored ($\Delta$BIC = 45.1). Notably, we found that the bimodality in the color distribution is discernible throughout the entire magnitude range covered by the Sloan observations, which counters the observation in \citet{gilhutton} of an apparent lack of low spectral slope objects in the range $10 < H < 12$. 

The bimodality in the color distribution indicates that the Hilda population is comprised of two types of objects, with characteristically different surface colors. Following the terminology in \citet{wong}, we refer to these as the less-red (LR) and red (R) Hildas. We did not detect any significant correlations between spectral slope and any orbital parameter, which demonstrates that LR and R Hildas are well-mixed within the overall population. 

We also studied the color distribution of Hilda and Schubart family members. The bottom panel of Figure~\ref{Hildacol} shows the color distribution of the two families along with the color distribution of non-family Hildas (i.e., background objects). The key observation here is that the color distribution of family members is unimodal and centered at relatively low spectral slope values consistent with the LR sub-population, whereas the background population (and the Hilda population overall) is bimodal in color. Examining the distribution of background objects with low spectral slope values in orbital parameter space, we do not find any notable correlation with the location of known family members; therefore, we do not expect significant contamination of collisional family members within the background population. Conversely, the handful of high spectral slope family members are likely interlopers and not formally collisional fragments. 

These results may indicate that the progenitor bodies of the Hilda and Schubart families were LR objects. Alternatively, the unimodal color distribution of collisional fragments may demonstrate the pristine interior material of Hilda asteroids, upon irradiation and space weathering, evolve to take on a less-red color, regardless of the color of the progenitor body. The latter possibility has important implications for the our understanding of the origin of the color bimodality, as we discuss in the next section.

\begin{figure}[t!]
\begin{center}
\includegraphics[width=9cm]{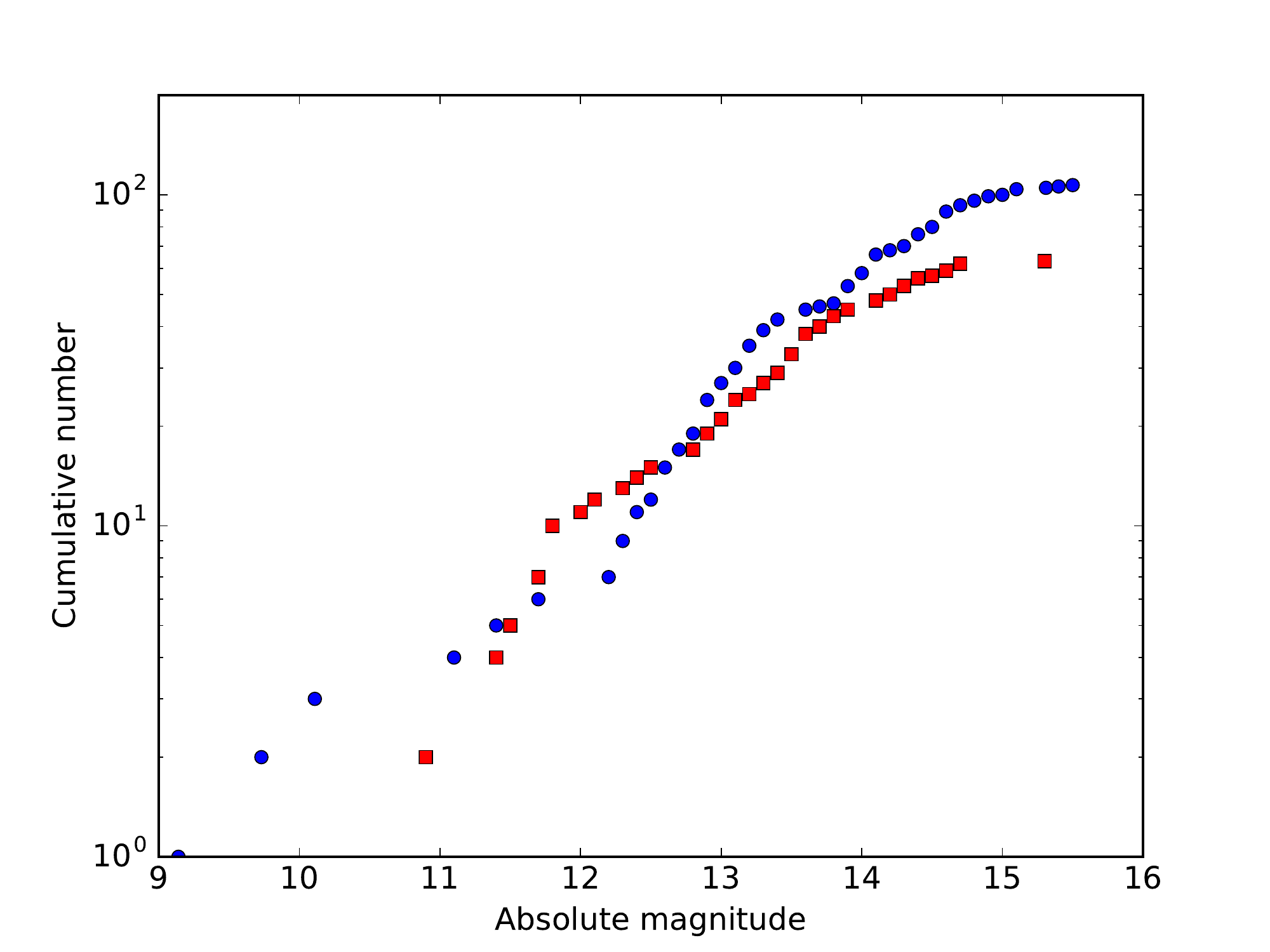}
\end{center}
\caption{The cumulative magnitude distributions of the LR and R sub-populations, where objects (including family members) have been categorized into the sub-populations by spectral slope. The distributions are statistically distinct from each other at the 98\% confidence level. Both distributions have a characteristically wavy shape that is not consistent with a single or double power law curve.} \label{Hildacolmag}
\end{figure}

In order to classify individual objects as LR or R Hildas, we fit the spectral slope distribution of background Hildas with a double Gaussian and obtained the mean colors of the LR and R sub-populations ---  $4.0 \times 10^{-5}~\mathrm{\AA}^{-1}$ and  $9.3 \times 10^{-5}~\mathrm{\AA}^{-1}$, respectively. We chose to remove family members in our calculation of mean colors since the surface composition of fragments may be systematically different than the surfaces of uncollided Hildas and would therefore not accurately reflect the initial color distribution. Using an analogous methodology to the one described in \citet{wong}, we categorized all Hildas (including collisional fragments) with $S \le 4.0 \times 10^{-5}~\mathrm{\AA}^{-1}$ as LR and all Hildas with $S \ge 9.3 \times 10^{-5}~\mathrm{\AA}^{-1}$ as R, resulting in a sample of 107 LR and 63 R Hildas. The cumulative absolute magnitude distributions of the LR and R sub-populations are shown in Figure~\ref{Hildacolmag}. 

Both distributions are characterized by wavy shapes that are not well-described by a single or double power law; we do not present distribution fits for the individual color sub-population magnitude distributions in this paper. Nevertheless, we compared the LR and R magnitude distributions using the two-sample Anderson-Darling test. We reject the null hypothesis that the LR and R magnitude distributions are sampled from a single underlying distribution at the 0.8\% significance level. In other words, the two color magnitude distributions are statistically distinct at the 99.2\% confidence level.

\section{Discussion}
Having carried out an analysis of the color-magnitude distribution of Hildas analogous to the one presented for Trojans in \citet{wong}, we are now in a position to compare the two populations. Recent dynamical instability models of Solar System evolution describe a common progenitor population of minor bodies in the primordial trans-Neptunian region from which both Hildas and Trojans are sourced. It follows that, if the current paradigm of Solar System evolution is correct, there should be notable similarities between the observable properties of the two populations.

\begin{figure}[t!]
\begin{center}
\includegraphics[width=9cm]{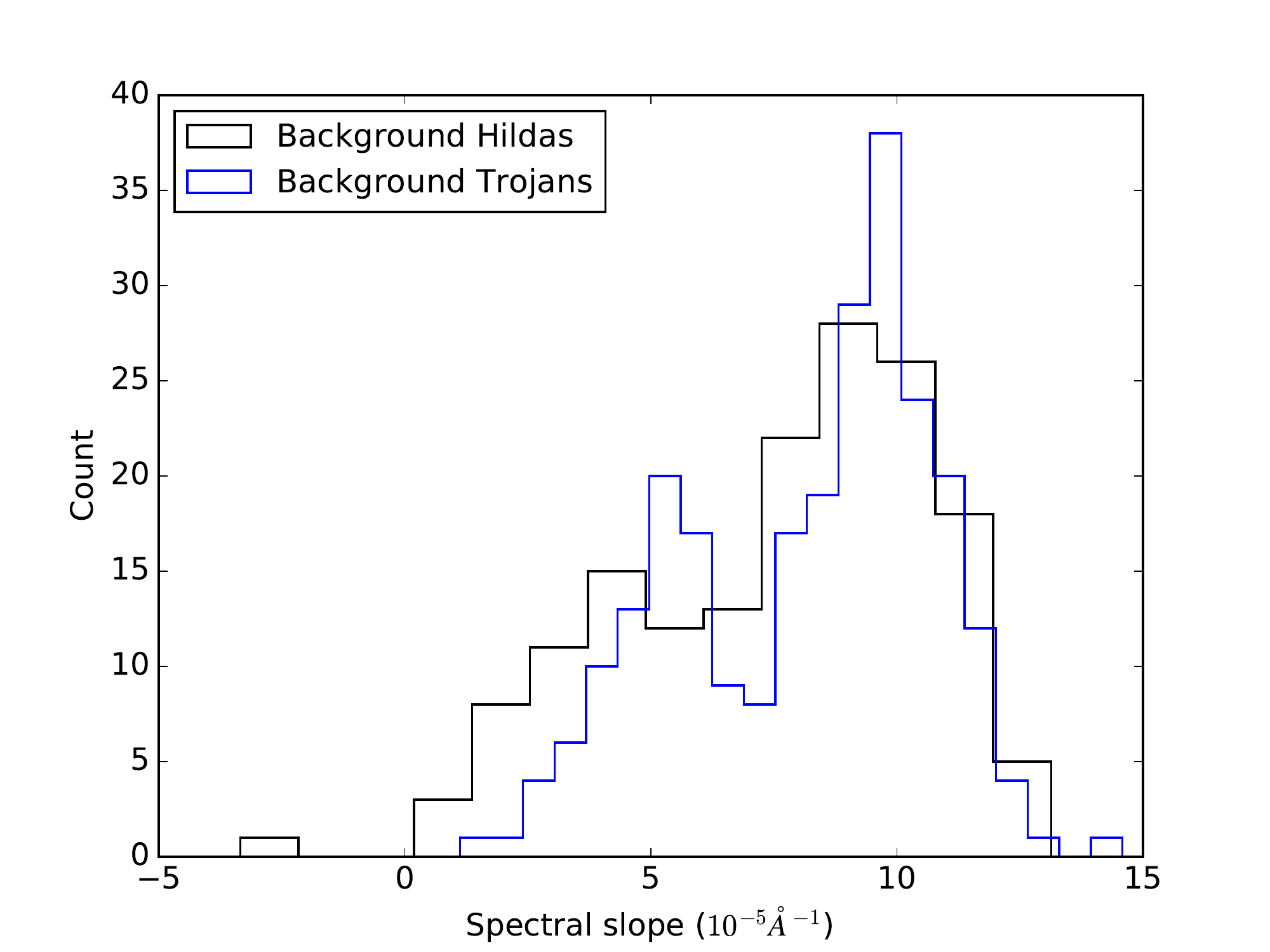}
\end{center}
\caption{Comparison of the Hilda and Trojan color distributions, with family members removed. For Hildas, all objects brighter than $H=14$ are shown, while for Trojans, all objects brighter than $H=12.3$ are shown; these are the established completeness limits of the corresponding analyses \citep[see][for the discussion of Trojans]{wong}. Both distributions show a clear bifurcation in color, corresponding to the LR and R sub-populations present in both populations, with comparable mean colors. The R-to-LR number ratio in both Hilda and Trojan background populations are also similar.} \label{colorcomp}
\end{figure}

The most salient similarity between Hildas and Trojans is their bimodal color distributions. Figure~\ref{colorcomp} shows the spectral slope distributions for Hildas and Trojans. Since fragments from a major collision introduce a significant number bias in the color distribution of a population relative to the initial pre-collision state, we have removed the Hilda and Schubart family members in order to compare the background populations only. Running the normal mixture model test on the background Hilda and Trojan color distributions (Section~2.2), we find that a two-peaked model is very strongly favored over a single-peaked model in both cases ($\Delta$BIC values of 16.7 and 39.1, respectively). From the figure, we can see that the characteristic mean colors of the LR and R Hildas and Trojans are comparable. The mean colors of the Trojan LR and R sub-populations are $5.3 \times 10^{-5}~\mathrm{\AA}^{-1}$ and  $9.6 \times 10^{-5}~\mathrm{\AA}^{-1}$, respectively, as compared to the somewhat bluer mean colors of the Hilda LR and R sub-populations ($4.0 \times 10^{-5}~\mathrm{\AA}^{-1}$ and  $9.3 \times 10^{-5}~\mathrm{\AA}^{-1}$, respectively). 

In addition, the number ratio of R-to-LR objects is similar for the Hildas and Trojans. In \citet{wong}, we categorized 47 background Trojans as LR and 104 as R, while in the present work, we obtained a categorized sample of 28 LR and 56 R objects out of the background Hilda population. For both non-family Hildas and Trojans, the R-to-LR number ratio is roughly 2-to-1. 

The origin of the color bimodality in the Trojans and Hildas has long remained unexplained. Earlier explanations concerning the color bimodality in Trojans suggested that the LR and R populations may have been sourced from different regions of the solar nebula, with one population originating in the middle Solar System and the other scattered in from the outer Solar System. However, within the framework of current dynamical instability models, such a scenario is not supported; instead, both LR and R Hildas and Trojans are predicted to have been emplaced from the same primordial collection of planetesimals in the outer Solar System.

In a hypothesis first proposed in \citet{wong} and subsequently developed in full in \citet{wong3}, we posited that the color bimodality arose within the primordial trans-Neptunian planetesimal disk, i.e., the purported progenitor population of Trojans and Hildas. In short, objects in this region accumulated out of a mix of rocky material and ices of roughly cometary composition, including a significant volume of volatile ices such as ammonia and methanol. Under the action of solar irradiation, location-dependent volatile loss led to differential surface depletion of the various volatile ices: objects closer in experienced higher surface temperatures and faster rates of sublimation, leading to the depletion of the more volatile species from the surface layers, while objects farther out were colder and thereby retained some of the more volatile species.

In the volatile loss model we developed, it was shown that H$_{2}$S would have been a key distinguishing factor, dividing the trans-Neptunian planetesimal population into two groups, with the closer objects depleted in H$_{2}$S on their surfaces and farther objects retaining H$_{2}$S. Irradiation of the volatile ice rich surfaces would have reddened and darkened the surfaces of all objects in the region, as has been demonstrated in various laboratory experiments \citep[e.g.,][]{brunetto}; however, irradiation of objects that retained H$_{2}$S on their surfaces would have produced sulfur-bearing molecules in the irradiated mantle, which is expected to provide a significant additional reddening \citep[e.g., as in the polar deposits on Io;][]{carlson}. As a result, we posited that the H$_{2}$S-retaining objects would have attained characteristically redder surface colors than the H$_{2}$S-depleted objects. The subsequent scattering of the trans-Neptunian planetesimal disk and the emplacement of Hildas and Trojans into their present-day locations would have preserved this primordial color bifurcation.

This hypothesis for the observed color bimodality has an important implication that explains another point of similarity between Hildas and Trojans --- the observation that all collisional family members are LR. As shown in Section~2.2, both the Hilda and Schubart families are comprised of exclusively LR objects; analyzed spectra of objects from the only robustly attested major collisional family in the Trojans --- the Eurybates family \citep{broz} --- reveal a similar pattern in which the family members have a unimodal color distribution centered at relatively low spectral slopes \citep{fornasier}. Regardless of the original surface color of the parent bodies, our color bimodality hypothesis offers a natural explanation for the observed trend. Upon a shattering impact, the fragments are composed of the pristine interior material of the parent bodies, namely, rocky material, water ice, and any remaining volatile ices trapped in the subsurface. At the much higher temperatures of the Hilda and Trojan regions, the volatile ices sublimate instantaneously from the surfaces of the family members. Consequently, irradiation of these volatile-depleted surfaces would not lead to reddening of the same extent as in the case where volatile ices are retained (in particular, H$_{2}$S), resulting in LR surface colors.

\begin{figure}[t!]
\begin{center}
\includegraphics[width=9cm]{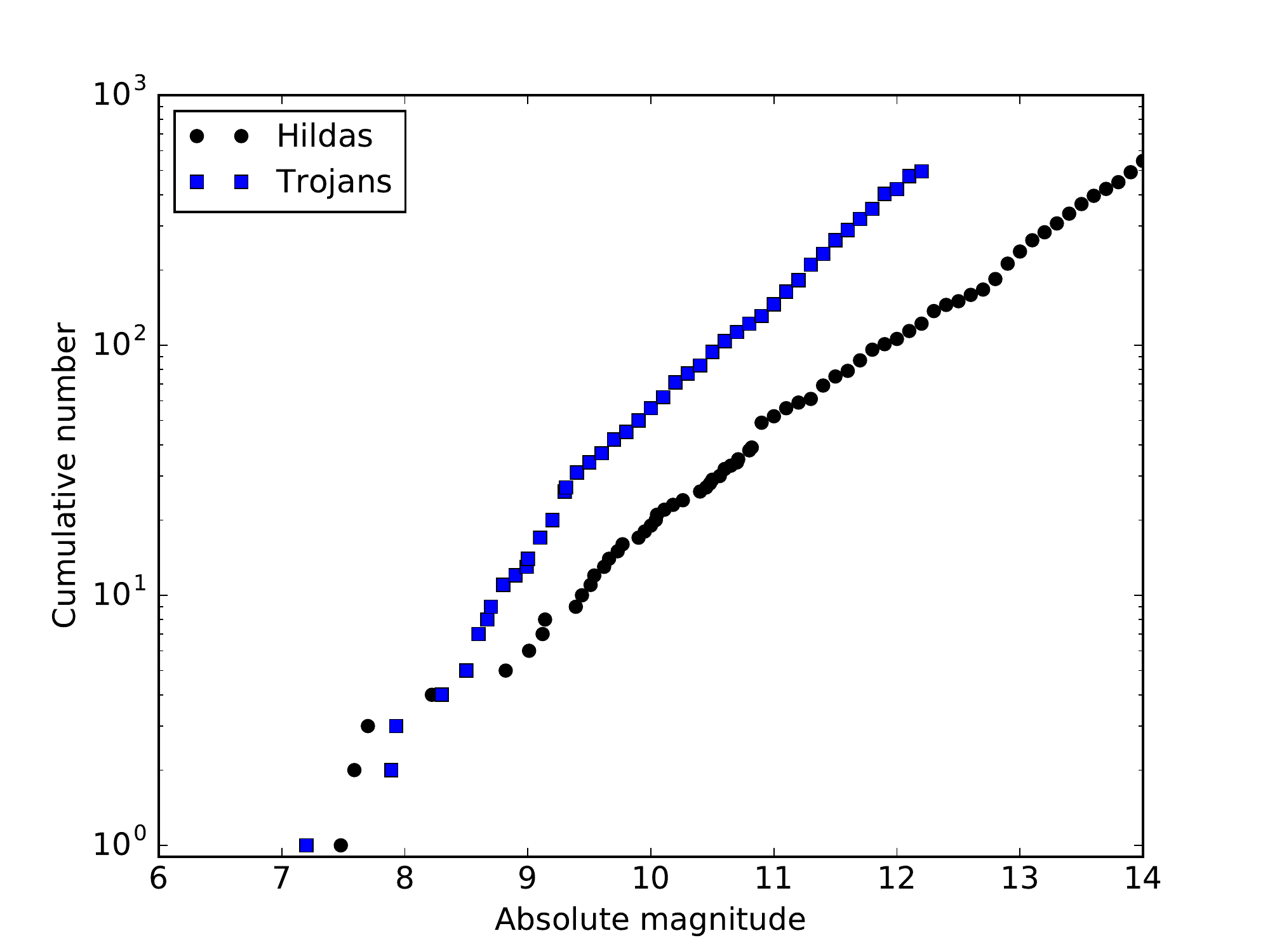}
\end{center}
\caption{Comparison of the total cumulative magnitude distributions of Hildas (black dots) and Trojans (blue squares). The Trojan distribution has been corrected for incompleteness, following the methods of \citet{wong}. The Hilda magnitude distribution is notably shallower throughout the entire magnitude range of the data.} \label{magcompare}
\end{figure}

Moving on to the magnitude distributions, we compare the total cumulative magnitude distributions for Hildas and Trojans in Figure~\ref{magcompare}. The general shape of the distributions is the same. In \citet{wong}, we modeled the collisional evolution of Trojans using the intrinsic collisional probabilities and impact velocities derived from previously published numerical simulations. We found that current level of collisional activity is insufficient to have produced the observed break in the magnitude distribution at $H\sim 9$ starting from a single power law initial magnitude distribution \citep[see also][]{marzari}; instead, the break is likely to be a consequence of the much more intense collisional environment in the early trans-Neptunian planetesimal region from which the Trojans and Hildas originated. From our modeling, we showed that the collisional evolution of the Trojans, assuming current rates, would only have resulted in a slight flattening of the power law slope at intermediate sizes.

A major difference between the Hilda and Trojan magnitude distributions is that the former is significantly shallower at all sizes. In particular, the power law slopes at intermediate sizes  --- $0.46 \pm 0.01$ \citep{wong} and $0.34^{+0.02}_{-0.01}$ (Section~2.1) for the Trojans and Hildas, respectively --- are discrepant at the $5.4\sigma$ level. In the context of collisional evolution and assuming that both Hildas and Trojans were derived from the same progenitor population and therefore were emplaced with similar initial size distributions, the shallower Hilda magnitude distribution would be indicative of a more active collisional environment. However, estimates of both the current intrinsic collisional probability and impact velocities are significantly lower for the Hildas than for the Trojans \citep[][and references therein]{davis}. This inconsistency presents a challenge to the idea of a common origin for the Hildas and Trojans as proposed by current dynamical instability models.

One possible explanation is apparent when considering the number of major collisional families in the Hildas and Trojans. Despite its lower current rate of collisional activity, the Hilda population contains two major collisional families, with the Hilda family containing the largest object (153 Hilda, $H=7.48$) in the entire population. Meanwhile, the Trojan population has only one major family, the Eurybates family, with its largest object (3548 Eurybates, $H=9.7$) being significantly smaller than 153 Hilda. All else being equal, the frequency of shattering collisions decreases sharply with increasing target size, due to the decrease in the number of impactors capable of fragmenting the target body. Assuming that the characteristic impact velocity of a resonant population does not change appreciably with time, one way of increasing the collisional probability for large targets is by increasing the number of impactors.

Therefore, the presence of two major collisional families in the Hildas suggests that perhaps the number of objects emplaced into the 3:2 resonance initially was much higher, creating a significantly more active early collisional environment, but was gradually depleted as collisional activity pushed fragments out of the resonance and out of the Hilda population. The 3:2 mean motion resonance with Jupiter has a narrow 0.1~AU-wide stable zone centered at 3.96~AU, surrounded on both sides by a dynamically chaotic boundary region with very short characteristic diffusion times \citep{ferraz}. In Figure~\ref{distall}, the location of the stable zone is evident in the sharp decrease in object density outside of the central region. 

Following a scenario that has been described by several earlier works \citep[e.g.,][]{gilhutton}, if a collisional fragment is ejected from the central stable region, it is removed from the resonance on a relatively short timescale, thereby depleting the magnitude distribution. The relative ejection velocity required for a fragment to pass out of the center of the resonance is around $\Delta V \sim 0.16$~km/s \citep{davis}. Based on the results of numerical models simulating the fragmentation of asteroidal bodies and given the characteristic impact velocity in the Hilda population \citep{davis}, a significant fraction of collisional fragments is expected to have a sufficient ejection velocity to exit the resonance \citep{jutzi}. These same simulations demonstrate that the smallest fragments tend to be imparted the highest ejection velocity.

Since the smaller bodies experience more frequent collisions and are also more likely to be ejected from the stable zone, the initial Hilda magnitude distribution would have become depleted most severely at faint magnitudes, consistent with the relatively shallow Hilda distribution when compared with the Trojan distribution. Eventually, as the total number of Hildas fell due to the removal of objects from resonance, the intrinsic collisional probability decreased to the present-day value.

Relating back to the important implication of our color bimodality hypothesis that collisional fragments are LR, a higher initial level of collisional activity also explains the distinct shapes of the LR and R Hilda magnitude distributions (Figure~\ref{Hildacolmag}). Collisions enrich the LR population exclusively and lead to a relative steepening in the shape of the LR magnitude distribution with time. The overall R-to-LR number should decrease with decreasing size, since collisions are much more frequent for smaller targets. In the Trojan population, the cumulative LR magnitude distribution was shown in \citet{wong2} to overtake the R magnitude distribution at $H\sim 15$. While the initial R-to-LR ratio of the Hildas was similar to that of the Trojans, as demonstrated by our earlier comparison of the background, uncollided color distribution (Figure~\ref{colorcomp}), we see that the LR sub-population becomes more numerous than the R sub-population at a larger size ($H\sim 13$), due to the significant enrichment of the LR sub-population by Hilda and Schubart family members.

All in all, the comparison of the magnitude and color distributions of Hildas and Trojans reveals several notable similarities, with the discrepancies in the present-day magnitude distribution shapes addressed by a dynamically plausible explanation. In turn, the body of observational data analyzed in this work presents a convincing case that the Hildas and Trojans originated from the same progenitor population prior to being emplaced in their current locations, as is predicted by current dynamical instability models of Solar System evolution. 

 \section{Conclusion}
In this paper, we analyzed the absolute magnitude and optical color distributions of the Hilda asteroids. We computed a power law fit to the magnitude distribution through $H=14$ and found a slope of $\alpha=0.34^{+0.02}_{-0.01}$.  Using photometric measurements contained in SDSS-MOC4, we calculated the spectral slope of 275 Hildas and confirmed the robust bimodality in color reported in \citet{gilhutton}. This bimodality demonstrates that the Hilda population is comprised of two groups of objects --- less-red and red Hildas. We classified individual objects into the two color sub-populations and presented the individual color magnitude distributions, which were shown to be highly distinct from each other. We also analyzed the Hilda and Schubart collisional families separately and found that the families are comprised of LR objects only.

Our comparison of the Hilda and Trojan color distributions revealed that both are bimodal, with similar characteristic LR and R colors. Furthermore, the R-to-LR number ratios among non-family Hildas and Trojans are consistent with each other; likewise, both populations display the same trend in which collisional family members are exclusively LR. Within the framework of dynamical instability models, our analysis of the Hilda and Trojan color distributions supports our previously published hypothesis that the color bimodality seen in both populations developed prior to emplacement in their current-day locations, with the difference in color primarily arising due to the retention vs. depletion of H$_{2}$S ice on the surfaces of planetesimals within the primordial trans-Neptunian disk.

Comparing the Hilda and Trojan total magnitude distributions, we showed that the Hilda distribution is significantly shallower than the Trojan distribution, despite being much less collisionally active at the present time. Upon consideration of the number of major collisional families in each population, we proposed an explanation for the discrepancy in magnitude distributions by positing that the Hilda population upon emplacement was significantly larger than the current population. This hypothesis naturally explains the higher apparent level of collisional evolution in the Hilda magnitude distributions (evidenced by the shallower power law slope at intermediate sizes), since small collisional fragments are readily ejected from the narrow stable zone of the 3:2 resonance and removed from the population. 

We conclude that the bulk properties of Hildas and Trojans lend strong support to the idea of a shared progenitor population --- a major step in validating one of the main predictions of current dynamical instability models of Solar System evolution. In further validation of these models, our photometric survey observations of small dynamically excited Kuiper Belt objects (KBOs) in the same size range as Hilda and Trojan asteroids reveal that these KBOs are likewise bimodal in optical color \citep{wong4}. We also show that the two color classes among the small KBOs have magnitude distributions that are statistically indistinguishable from the magnitude distributions of LR and R Trojans. Taken together, these studies provide the first body of observational evidence linking the properties of KBOs, Hildas, and Trojans. 

The question of the composition of HIldas, Trojans, and similarly sized KBOs remains unresolved and could provide a complementary probe into the similarities and/or differences between the respective asteroid populations. Intensive spectroscopic observations of Hildas and Trojans using current and near-future instruments promise to provide improved constraints on the surface composition of these bodies, which will help solidify our understanding of their origin.

\end{document}